\documentclass[graybox,envcountsect]{svmult}

\usepackage{mathptmx}       
\usepackage{helvet}         
\usepackage{courier}        
\usepackage{type1cm}        
\usepackage{graphicx}        
\usepackage[bottom]{footmisc}

\usepackage{amsmath,amssymb,latexsym}
\usepackage{url}

\def\wh{\widehat}

\def\A{\mbox{$\mathcal A$}}
\def\X{\mbox{$\mathcal X$}}

\def\F{\mbox{$\mathcal F$}}

\def\B{\mbox{$\mathcal B$}}
\def\C{\mbox{$\mathcal C$}}

\def\R{\mbox{$\mathcal R$}}
\def\T{\mbox{$\mathcal T$}}
\def\H{\mbox{$\mathcal H$}}

\def\FF{\mathbb F}
\def\Abb{\mathbb A}
\def\GG{\mathbb G}
\def\KK{\mathbb K}
\def\CC{\mathbb C}
\def\BB{\mathbb B}
\def\DD{\mathbb D}
\def\ZZ{\mathbb Z}

\def\RR{\mathbb R}
\def\EE{\mathbb{E}}
\def\PP{\mathbb{P}}
\def\GG{\mathbb{G}}

\def\YY{\mathbb{Y}}

\def\Xf{\mathbf{X}}
\def\Uf{\mathbf{U}}
\def\Vf{\mathbf{V}}
\def\xf{\mathbf{x}}
\def\yf{\mathbf{y}}
\def\uf{\mathbf{u}}
\def\vf{\mathbf{v}}
\def\tf{\mathbf{t}}
\def\1{\mathbf{1}}

\begin{document}

\title{An overview of the goodness-of-fit test problem for copulas.}
\titlerunning{Copula GOF tests}
\author{Jean-David Fermanian}
\institute{Jean-David Fermanian \at Department of Finance,
Ensae-Crest, Malakoff,
France,~\email{jean-david.fermanian@ensae.fr}}

\maketitle

\abstract{We review the main ``omnibus procedures'' for
goodness-of-fit testing for copulas: tests based on the empirical
copula process, on probability integral transformations, on
Kendall's dependence function, etc, and some corresponding
reductions of dimension techniques. The problems of finding
asymptotic distribution-free test statistics and the calculation of
reliable $p$-values are discussed. Some particular cases, like
convenient tests for time-dependent copulas, for Archimedean or extreme-value copulas, etc, are dealt with.
Finally, the practical performances of the proposed approaches
are briefly summarized.}

\medskip

\section{Introduction}\label{JDF_sec:1}

Once a model has been stated and estimated, a key question is to
check whether the initial model assumptions are realistic. In other
words, and even it is sometimes eluted, every modeler is faced with
the so-called ``goodness-of-fit'' (GOF) problem. This is an
old-dated statistical problem, that can be rewritten as: denoting by
$F$ the cumulative distribution function (cdf hereafter) of every
observation, we would like to test
$$\H_0: F=F_0,\;\; \text{against} \;\;\H_a: F\neq F_0,$$
for a given cdf $F_0$, or, more commonly,
$$\H_0: F\in \F,\;\; \text{against} \;\;\H_a: F\not\in \F,$$
for a given family of distributions $\F:=\{ F_\theta,\theta\in
\Theta\}$.
This distinction between simple and composite assumptions is traditional and we keep it.
Nonetheless, except in some particular cases (test of independence, e.g.), the latter framework is a lot more useful than the former in practice.

\medskip

Some testing procedures are ``universal'' (or ``omnibus''), in the sense
they can be applied whatever the underlying distribution. In other
terms, they do not depend on some particular properties of $F_0$ or
of the assumed family $\F$. Such tests are of primary interest for
us. Note that we will not consider Bayesian testing procedures, as
proposed in~\cite{JDF_Huard}, for instance.

\medskip

To fix the ideas, consider an i.i.d. sample $(\Xf_1,\ldots,\Xf_n)$
of a $d$-dimensional random vector $\Xf$. Its joint cdf is denoted
by $F$, and the associated marginal cdfs' by $F_j$, $j=1,\ldots,d$.
Traditional key quantities are provided by the empirical
distribution functions of the previous sample: for every $\xf\in
\RR^d$, set $d$ marginal cdfs'
$$F_{n,k}(x_k):=n^{-1}\sum_{i=1}^n \1 (X_{i,k} \leq x_k),\; k=1,\ldots,d,$$
and the joint empirical cdf $F_n(\xf):=n^{-1}\sum_{i=1}^n \1 (\Xf_i
\leq \xf).$ The latter inequality has to be understood
componentwise. Most of the ``omnibus'' tests are
based on transformations of the underlying empirical distribution
function, or of the empirical process $\FF_n:=\sqrt{n}(F_n-F_0)$
itself: $T_n = \psi_n(F_n)$ or $T_n = \psi_n(\FF_n)$. It is the case
of the famous Kolmogorov-Smirnov (KS), Anderson-Darling (AD),
Cramer-von-Mises (CvM) and chi-squared tests, for example.

\medskip

Naively, it could be thought the picture is the same for copulas,
and that straightforward modifications of standard GOF tests should
do the job. Indeed, the problem for copulas can be simply written as
testing
$$\H_0: C=C_0,\;\; \text{against}\;\; \H_a: C\neq C_0,$$
$$\H_0: C\in \C,\;\; \text{against}\;\; \H_a: C\not\in \C,$$
for some copula family $\C:=\{ C_\theta,\theta\in \Theta\}$.
Moreover, empirical copulas, introduced by Deheuvels in the 80's
(see~\cite{JDF_Deheu1978},~\cite{JDF_Deheu1981a},~\cite{JDF_Deheu1981b})
play the same role for copulas as standard empirical cdfs' for
general distributions. For any $\uf\in [0,1]^d$, they can be defined by
$$C_n(\uf):=F_n(F_{n,1}^{(-1)}(u_1),\ldots,F_{n,d}^{(-1)}(u_d)),$$
with the help of generalized inverse functions, or by
$$ \bar C_n(\uf):= \frac{1}{n}\sum_{i=1}^n \1 ( F_{n,1}(X_{i,1})\leq u_1,\ldots,F_{n,d}(X_{i,d})\leq u_d).$$
It can be proved easily that $\| C_n - \bar C_n \|_{\infty}\leq
dn^{-1}$ (see~\cite{JDF_FermRaduWeg1}). Then, for the purpose of GOF
testing, working with $C_n$ or $\bar C_n$ does not make any difference
asymptotically. In every case, empirical copulas are explicit
functionals of the underlying empirical cdf: $C_n = \zeta(F_n)$.
Thus, any previous GOF test statistics for copulas could be defined
as $T_n= \psi_n(C_n)=\psi_n\circ \zeta(F_n)$. But this functional
$\zeta$ is sufficient to induce significant technical difficulties,
when applied to standard statistical procedures.

\medskip

Actually, the latter parallel applies formally, but strong
differences appear in terms of the limiting laws of the
``copula-related'' GOF test statistics. Indeed, some of them are
distribution-free in the standard case, i.e., their limiting laws
under the null do not depend on the true underlying law $F$, and
then, they can be tabulated: KS (in the univariate case),
chi-squared tests, for example. Unfortunately, it is almost
impossible to get such nice results for copulas, due to their
multivariate nature and due to the complexity of the previous
mapping between $F_n$ and $C_n$. Only a few GOF test techniques for
copulas induce distribution-free limiting laws. Therefore, most of the time, some
simulation-based procedures have been proposed for this task.

\medskip

In section~\ref{JDF_sec:2}, we discuss the ``brute-force''
approaches based on some distances between the empirical copula
$C_n$ and the assumed copula (under the null), and we review the
associated bootstrap-like techniques. We detail how to get
asymptotically distribution-free test statistics in
section~\ref{JDF_sec:3}, and we explain some testing procedures that
exploit the particular features of copulas. We discuss some ways of
testing the belonging to some ``large'' infinite-dimensional
families of copulas like Archimedean, extreme-value, vine, or HAC copulas in
section~\ref{JDF_sec:4}. Tests adapted to time-dependent copulas are
introduced in section~\ref{JDF_sec:5}. Finally, empirical
performances of these GOF tests are discussed in
section~\ref{JDF_sec:6}.

\medskip

\section{The ``brute-force'' approach: the empirical copula process and the bootstrap}\label{JDF_sec:2}

\subsection{Some tests based on empirical copula
processes}
\label{JDF_subsec:1}

Such copula GOF tests are the parallels of the most standard GOF
tests in the literature, replacing $F_n$ (resp. $F_0$) by $C_n$
(resp. $C_0$). These statistics are based on distances between the
empirical copula $C_n$ and the true copula $C_0$ (simple zero
assumption), or between $C_n$ and $C_{\hat\theta_n}$ (composite zero
assumption), for some convergent and convenient estimator $\hat\theta_n$ of the ``true''
copula parameter $\theta_0$. It is often reduced simply to the evaluation of
norms of the empirical copula process $\CC_n :=\sqrt{n}(C_n-C_0)$,
or one of its approximations $\widehat \CC_n
:=\sqrt{n}(C_n-C_{\hat\theta_n})$.

\medskip

In this family, let us cite the Kolmogorov-Smirnov type statistics
$$T^{KS}_n :=\| \CC_n
\|_\infty= \sup_{\uf \in [0,1]^d}|\sqrt{n}(C_n-C_0)(\uf)|,$$ and the
Anderson-Darling type statistics
$$T^{AD}_n :=\| \CC_n
\|_{L^2}= n\int (C_n-C_0)^2(\uf)w_n(\uf)\, d\uf,$$ for some positive
(possibly random) weight function $w_n$, and their composite versions. By smoothing
conveniently the empirical copula process,~\cite{JDF_Omelka} defined
alternative versions of the latter tests.

\medskip

In practice, the statistics $T^{KS}_n$ seem to be less powerful than
a lot of competitors, particularly of the type $T^{AD}_n$
(see~\cite{JDF_BeauGenestRem}). Therefore, a ``total variation''
version of $T^{KS}_n$ has been proposed in~\cite{JDF_FermRaduWeg2},
that appears significantly more powerful than the classical
$T_n^{KS}$:
$$T_{n}^{ATV}:=\sup_{B_{1},\ldots
,B_{L_n}}\sum_{k=1}^{L_n}|\CC_{n}(B_{k})|,\;\;\text{or}\;\; \hat
T_{n}^{ATV}:=\sup_{B_{1},\ldots
,B_{L_n}}\sum_{k=1}^{L_n}|\widehat{\CC}_{n}(B_{k})|,$$ for simple or
composite assumptions respectively. Above, the supremum is taken
over all disjoint rectangles $B_{1},\ldots ,B_{L_n}\subset \lbrack
0,1]^{d}$, and $L_n\sim \ln n$.

\medskip

Another example of distance is proposed in~\cite{JDF_Panch}: let
two functions $f_1$ and $f_2$ in $\RR^d$. Typically, they
represent copula densities. Set a positive definite bilinear
form as
$$<f_1,f_2>:=\int \kappa_d(\xf_1,\xf_2)\,
f_1(\xf_1)f_2(\xf_2)\, d\xf_1\, d\xf_2,$$ where
$\kappa_d(\xf_1,\xf_2):=\exp( - \| \xf_1 - \xf_2 \|^2/ (2dh^2))$,
for some Euclidian norm $\|\cdot \|$ in $\RR^d$ and a bandwidth
$h>0$. A squared distance between $f_1$ and $f_2$ is given simply by
$\mu(f_1,f_2):=<f_1-f_2,f_1-f_2>=<f_1,f_1> - 2 <f_1,f_2> + <f_2,f_2>.$ When $f_1$ and
$f_2$ are the copula densities of $C_1$ and $C_2$ respectively,
the three latter terms can be rewritten in terms of copula directly. For
instance, $<f_1,f_2>= \int \kappa_d(\xf_1,\xf_2)\, \, C_1(d\xf_1)\,
C_2(d\xf_2).$ Since such expressions have simple empirical
counterparts, a GOF test for copulas can be built easily: typically,
replace $C_1$ by the empirical copula $C_n$ and $C_2$ by the true
copula $C_0$ (or $C_{\hat\theta_n}$).

\medskip

Closely connected to this family of tests are statistics $T_n$
that are zero when the associated copula processes are zero, but
not the opposite. Strictly speaking, this is the case of the
Cramer-von Mises statistics
$$T^{CvM}_n :=n\int (C_n-C_0)^2(\uf)\, C_n(d\uf),$$
and of chi-squared type test statistics, like
$$T^{Chi}_n := n\sum_{k=1}^p w_k(C_n-C_0)^2(B_k) ,$$
where $B_1,\ldots,B_p$ denote disjoint boxes in $[0,1]^d$ and
$w_k$, $k=1,\ldots,p$ are convenient weights (possibly random).
More generally, we can consider
$$T^{\mu}_n := \sum_{k=1}^p \mu(C_n(E_k),C_0(E_k)),\;\;\text{or}\;\; T^{\mu}_n :=
\sum_{k=1}^p \mu(C_n(E_k),C_{\hat \theta_n}(E_k)), $$ for any metric
$\mu$ on the real line, and arbitrary subsets $E_1,\ldots,E_p$ in
$[0,1]^d$. This is the idea of the chi-square test detailed
in~\cite{JDF_DobricSchmid2005}: set the vectors of pseudo-observations
$\hat\Uf_{i}:=(F_{n,1}(X_{i,1}),\ldots, F_{n,d}(X_{i,d}))$, and a
partition of $[0,1]^d$ into $p$ disjoint rectangles $B_j$. The
natural chi-square-style test statistics is
$$ T_n^\chi:=\sum_{k=1}^p \frac{\left( \hat N_k - p_k(\hat\theta_n)
\right)^2}{np_k(\hat\theta_n)}$$ where $\hat N_k$ denotes the number
of vectors $\hat\Uf_{i}$, $i=1,\ldots,n$ that belong to $B_k$, and
$p_k(\theta)$ denotes the probability of the event $\{ \Uf \in
B_k\}$ under the copula $C_\theta$. This idea of applying an
arbitrary categorization of the data into contingency tables
$[0,1]^d$ has been applied more or less fruitfully in a lot of
papers:~\cite{JDF_GenestRivest},~\cite{JDF_Klugman},~\cite{JDF_Ferm},~\cite{JDF_Andersen},~\cite{JDF_JunkerMay},
etc.

\medskip

Finally, note that a likelihood ratio test has been proposed
in~\cite{JDF_DobricSchmid2005}, based on a Kullback-Leibler pseudo
distance between a "discrete" version of $C_n$ and the corresponding
estimated copula under the null:
$$ T_n^{LR}:= \sum_{k=1}^p N_k \ln p_k(\hat\theta_n).$$
To compare the fit of two potential parametric copulas, the same
information criterion has been used in~\cite{JDF_Diks} to build a
similar test statistics, but based on copula densities directly.

\medskip

The convergence of all these tests relies crucially on the fact
that the empirical copula processes $\CC_n$ and $\widehat\CC_n$
are weakly convergent under the null, and for convenient sequences
of estimates $\hat\theta_n$:
see~\cite{JDF_Ruschendorf},~\cite{JDF_GaenslerStute},~\cite{JDF_FermRaduWeg1}.
Particularly, it has been proved that $\CC_n$ tends weakly in
$\ell^\infty([0,1]^d)$ (equipped with the metric induced by the
sup-norm) to a Gaussian process $\GG_{C_0}$, where
$$ \GG_{C_0}(\uf):= \BB_{C_0}(\uf) - \sum_{j=1}^d \partial_j C_0(\uf)
\BB_{C_0}(u_j,\1_{-j}),\; \forall \uf \in [0,1]^d,$$ with obvious
notations and for some $d$-dimensional Brownian bridge $\BB$ in
$[0,1]^d$, whose covariance is
$$\EE\left[\GG_{C_0}(\uf)\GG_{C_0}(\vf)   \right] = C_0(\uf\wedge \vf) - C_0(\uf)C_0(\vf),\;\;
\forall (\uf,\vf) \in [0,1]^{2d}.$$
To get this weak convergence result, it is not necessary to assume that $C_0$ is continuously differentiable on the whole hypercube $[0,1]^d$, a condition that is often not fulfilled in practice.
Recently,~\cite{JDF_Segers} has shown that such a result is true when, for every $j=1,\ldots,d$, $\partial_j C_0$ exists and is continuous on the set $\{\uf \in [0,1]^d, 0<u_j<1\}$.

\medskip

Clearly, the law of $\GG$ involves the particular
underlying copula $C_0$ strongly, contrary to usual
Brownian bridges. Therefore, the tabulation of the limiting laws of
$T_n$ GOF statistics appears difficult. A natural idea is to rely on
computer intensive methods to approximate these law numerically. The
bootstrap appeared as a natural tool for doing this task

\medskip

\subsection{Bootstrap techniques}\label{JDF_subsec:2}

The standard nonparametric bootstrap is based on resampling with
replacement inside an original i.i.d. $\Xf$-sample $S_\Xf$. We get
new samples $S_\Xf^*=(\Xf_1^*,\ldots,\Xf_n^*)$. Associate to every
new sample $S_\Xf^*$ its ``bootstrapped'' empirical copula
$C_n^*$ and its bootstrapped empirical process
$\CC_n^*:=\sqrt{n}(C_n^* - C_n)$. In~\cite{JDF_FermRaduWeg1}, it is
proved that, under mild conditions, this bootstrapped process
$\CC_n^*$ is weakly convergent in $\ell^\infty([0,1]^d)$ towards the
previous Gaussian process $\GG_{C_0}$. Therefore, in the case of
simple null assumptions, we can get easily some critical values or
p-values of the previous GOF tests: resample $M$ times, $M>>1$, and
calculate the empirical quantiles of the obtained bootstrapped test
statistics. Nonetheless, this task has to be done for every zero
assumption. This can become a tedious and rather long task,
especially when $d$ is ``large'' ($>3$ in practice) and/or with
large datasets ($>1000$, typically).

\medskip

When dealing with composite assumptions, some versions of the
parametric bootstrap are advocated, depending on the limiting
behavior of $\hat\theta_n - \theta_0$: see the theory in
~\cite{JDF_GenestRem}, and the appendices
in~\cite{JDF_BeauGenestRem} for detailed examples. To summarize
these ideas in typical cases, it is now necessary to draw random
samples from $C_{\hat\theta_n}$. For every bootstrapped sample,
calculate the associated empirical copula $C_n^*$ and a new
estimated value $\hat\theta_n^*$ of the parameter. Since the weak
limit of $\sqrt{n}(C_n^*-C_{\hat\theta_n^*})$ is the same as the
limit of $\widehat\CC_n=\sqrt{n}(C_n-C_{\hat\theta_n})$, the law of
every functional of $\widehat{\CC}_n$ can be approximated. When the cdf
$C_{\hat\theta_n}$ cannot be evaluated explicitly (in closed-form),
a two-level parametric bootstrap has been proposed
in~\cite{JDF_GenestRem}, by bootstrapping first a approximated
version of $C_{\hat\theta_n}$.

\medskip

Instead of resampling with replacement, a multiplier bootstrap
procedure can approximate the limiting process $\GG_{C_0}$
(or one of its functionals), as in~\cite{JDF_ScaillRem}: consider
$Z_1,\ldots,Z_n$ i.i.d. real centered random variables with variance
one, independent of the data $\Xf_1,\ldots,\Xf_n$. A new
bootstrapped empirical copula is defined by
$$ C_n^*(\uf):=\frac{1}{n}\sum_{i=1}^n Z_i .\1 (F_{n,1}(X_{i,1})\leq u_1,\ldots, F_{n,d}(X_{i,d})\leq
u_d),$$ for every $\uf\in [0,1]^d$. Setting $\bar
Z_n:=n^{-1}\sum_{i=1}^n Z_i$, the process $\beta_n:=\sqrt{n}(C_ n^*
- \bar Z_n C_n)$ tends weakly to the Brownian bridge $\BB_{C_0}$. By
approximating (by finite differences) the derivatives of the true
copula function, it is shown in~\cite{JDF_ScaillRem} how to modify
$\beta_n$ to get an approximation of $\GG_{C_0}$. To avoid this last
stage, another bootstrap procedure has been proposed
in~\cite{JDF_BucherDette2010}. It applies the multiplier idea to the
underlying joint and marginal cdfs', and invoke classical delta
method arguments. Nonetheless, despite more attractive theoretical
properties, the latter technique does not seem to improve the
initial multiplier bootstrap of~\cite{JDF_ScaillRem}.
In~\cite{JDF_KojadYan2011}, the multiplier approach is extended to
deal with parametric copula families of any dimension, and the
finite-sample performance of the associated Cramer-von-Mises test
statistics has been studied. A variant of the multiplier approach
has been proposed in~\cite{JDF_KojadYan2009}. It is shown that the
use of multiplier approaches instead of the parametric bootstrap
leads to a strong reduction in the computing time. Note that both methods have been implemented in the copula R package.

\medskip

Recently, in~\cite{JDF_FermRaduWeg2}, a modified nonparametric
bootstrap technique has been introduced to evaluate the limiting law
of the previous Komogorov-Smirnov type test statistics $T_n^{ATV}$
in the case of composite zero assumptions. In this case, the key
process is still
$$\widehat{\CC}_n := \sqrt{n}(C_n
-C_{\hat\theta_n})=\CC_n -\sqrt{n}(C_{\hat\theta_n}-C_{\theta_0}).$$
Generate a usual nonparametric bootstrap sample, obtained after
resampling with replacement from the original sample. This allows
the calculation of the bootstrapped empirical copula $C_n^*$ and a
new parameter estimate $\hat\theta_n^*$. Instead of considering the
``intuitive'' bootstrapped empirical copula process $\sqrt{n}(C^*_n
- C_{\hat\theta_n^*})$, a new bootstrapped process is introduced:
$$\YY_n^*:=\sqrt{n}(C_n^* - C_n) - \sqrt{n}(C_{\hat\theta_n^*} - C_{\hat\theta_n}).$$
Indeed, the  process $\sqrt{n}(C_n^*-C_{\wh\theta^*})$, while
perhaps a natural candidate, does not yield a consistent estimate of
the distribution of $\widehat\CC_n$, contrary to $\YY_n^*$. For
the moment, the performances of this new bootstrapped process have
to be studied more in depth.

\medskip

\section{Copula GOF test statistics: alternative approaches}\label{JDF_sec:3}

\subsection{Working with copula densities}\label{JDF_subsec:3}

Even if the limiting laws of the empirical copula processes
$\CC_n$ and $\widehat{\CC}_n$ involve the underlying (true) copula
in a rather complex way, it is still possible to get
asymptotically distribution-free test statistics. Unfortunately,
the price to be paid is an additional level of complexity.

\medskip

To the best of our knowledge, there exists a single strategy. The
idea is to rely on copula densities themselves, rather than copulas
(cdfs'). Indeed, testing the identity $C=C_0$ is equivalent to
studying the closeness between the true copula density $\tau_0$ (w.r.t.
the Lebesgue measure on $[0,1]^d$, that is assumed to exist) and one of its estimates $\tau_n$.
In~\cite{JDF_Ferm}, a $L^2$-distance between $\tau_n$ and $\tau_0$
allows to build convenient test statistics. To be specific, a kernel
estimator of a copula density $\tau$ at point $\uf$ is defined by
$$ \tau_n(\uf)=\frac{1}{h^d}\int K\left(\frac{ \uf - \vf }{h} \right)
C_n(d\vf)= \frac{1}{nh^d} \sum_{i=1}^n K\left(\frac{\uf- \hat\Uf_{i}
}{h} \right),$$
 where $\hat\Uf_{i}:=(F_{n,1}(X_{i,1}),\ldots,F_{n,d}(X_{i,d}))$ for all $i=1,\ldots,n$.
Moreover, $K$ is a $d$-dimensional kernel and $h=h(n)$ is a
bandwidth sequence, chosen conveniently. Under some regularity
assumptions, for every $m$ and every vectors $\uf_1,\ldots,\uf_m$
in $]0,1[^d$, such that $\tau_0(\uf_k)>0$ for every $k$, the vector
$ (nh^d)^{1/2} \left( (\tau_n - \tau_0)(\uf_1),\ldots,(\tau_n -
\tau_0)(\uf_m)\right)$ tends weakly to a Gaussian random vector,
whose components are independent. Therefore, under the null, the
test statistics
$$ T_n^{\tau,0}= \frac{nh^d}{\int K^2} \sum_{k=1}^m \frac{ (\tau_n(\uf_k) -
\tau_0(\uf_k) )^2 }{\tau_0(\uf_k)^2 }\cdot$$ tends
in law towards a $m$-dimensional chi-squared distribution. This can
be adapted easily for composite assumptions. The previous test
statistics depend on a finite and arbitrary set of points $\uf_k$,
$k=1,\ldots,m$. To avoid this drawback,~\cite{JDF_Ferm} has introduced
$$ J_n = \int (\tau_n - K_h\ast \hat \tau)^2(\uf)\omega(\uf)\, d\uf,$$
for some nonnegative weight function $\omega$. Here, $\hat\tau$
denotes $\tau_0$ (simple assumption) or $\tau(\cdot,\hat\theta_n)$
(composite assumption), for sufficiently regular estimates
$\hat\theta_n$ of $\theta_0$. It is proved that
$$ T_n^{\tau,1} := \frac{n^2h^{d}\left( J_n - (nh^d)^{-1}\int K^2(\tf).
(\hat\tau\omega)(\uf-h\tf) \,d\tf\, d\uf + (nh)^{-1}\int
\hat\tau^2\omega .\sum_{r=1}^d\int K_r^2 \right)^2 }{2
\int\hat\tau^2 \omega\cdot \int \left\{ \int K(\uf)K(\uf+\vf)\,
d\uf \right\}^2 \, d\vf }$$ tends to a $\chi^2(1)$ under the
null.

\medskip

Even if the previous test statistics are pivotal, they are rather
complex and require the choice of smoothing parameters and kernels.
Nonetheless, such ideas have been extended in~\cite{JDF_Scaill} to
deal with the fixed design case. Moreover, the properties of these
tests under fixed alternatives are studied
in~\cite{JDF_BucherDette2009}. The impact of several choices of
parameter estimates $\hat\theta_n$ on the asymptotic behavior of
$J_n$ is detailed too. Apparently, for small sample sizes, the
normal approximation does not provide sufficiently exact critical
values (in line with~\cite{JDF_HardleMammen}
or~\cite{JDF_FanLinton}), but it is still possible to use a
parametric bootstrap procedure to evaluate the limiting law of
$T_n^\tau$ in this case. Apparently, in the latter case, the results
are as good as the main competitors (see~\cite{JDF_BucherDette2009},
section 5).

\medskip

Since copula densities have a compact support, kernel smoothing can
generate some undesirable boundary effects. One solution is to use
improved kernel estimators that take care of the typical corner bias
problem, as in~\cite{JDF_Omelka}. Another solution is to estimate
copula densities through wavelets, for which the border effects are
handled automatically, due to the good localization properties of
the wavelet basis: see~\cite{JDF_GenestMasielloTribouley}. This idea
has been developed in~\cite{JDF_GayraudTribouley}, in a minimax
theory framework, to determine the largest alternative for which the
decision remains feasible. Here, the copula densities under
consideration are supposed to belong to a range of Besov balls.
According to the minimax approach, the testing problem is then
solved in an adaptive framework.

\medskip

\subsection{The probability integral transformation (PIT)}\label{JDF_subsec:4}

A rather simple result of probability theory, proposed initially
in~\cite{JDF_Rosenblatt}, has attracted the attention of authors for
copula GOF testing purpose. Indeed, this transformation maps a
general $d$-dimensional random vector $\Xf$ into a vector of $d$
independent uniform random variables on $[0,1]$ in a one-to-one
way. It is known as Rosenblatt's or probability integral
transformation (PIT). Once the joint law of $\Xf$ is known and
analytically tractable, this is a universal way of generating
independent and uniform random vectors without losing statistical
information. Note that other transformations of the same type exist
(see~\cite{JDF_Agostino}).

\medskip

To be specific, the copula $C$ is the joint cdf of
$\Uf:=(F_1(X_1),\ldots,F_d(X_d))$. We define the $d$-dimensional
random vector $\Vf$ by
\begin{equation}
V_1 := U_1=F_1(Z_1),\;V_2 := C(U_2 | U_1),\cdots,V_d:= C(U_d | U_1,\ldots,U_{d-1}), \label{JDF_PIT}
\end{equation}
where $C(\cdot|u_1,\ldots,u_{k-1})$ is the law of $U_k$ given
$U_1=u_1$,...,$U_{k-1}=u_{k-1}$, $k=2,\ldots,d$. Then, the variables
$V_k$, $k=1,\ldots,d$ are uniformly and independently distributed on
$[0,1]$. In other words, $\Uf \sim C$ iff $\Vf=\R(\Uf) $ follows the
$d$-variate independence copula $C_\perp(\uf)=u_1.\cdots.u_d$.

\medskip

The main advantage of this transformation is the simplicity of the
transformed vector $\Vf$. This implies that the zero assumptions of
a GOF test based on $\Vf$ are always the same: test the i.i.d.
feature of $\Vf$, that is satisfied when $C$ is the true underlying
copula. A drawback is the arbitrariness in the choice of the
successive margins. Indeed, there are at most $d!$ different PITs',
that induce generally different test statistics. Another disadvantage is the necessity of
potentially tedious calculations. Indeed, typically, the conditional
joint distributions are calculated through the formulas
$$C(u_k|u_1,\ldots,u_{k-1}) = \partial^{k-1}_{1,2,\ldots,k-1} C(u_1,\ldots,u_k,1,\ldots,1)/
\partial^{k-1}_{1,2,\ldots,k-1} C(u_1,\ldots,u_{k-1},1,\ldots,1),$$
for every $k=2,\ldots,d$ and every $\uf\in [0,1]^d$. Therefore, with
some copula families and/or with large dimensions $d$, the explicit
calculation (and coding!) of the PIT can become unfeasible.

\medskip

The application of such transformations for copula GOF testing
appeared first in~\cite{JDF_BreymannDias}. This idea has been
reworked and extended in several papers afterwards:
see~\cite{JDF_DobricSchmid2007},~\cite{JDF_BergBakken2007},~\cite{JDF_GenestQuessyRem2006a},~\cite{JDF_Berg},
etc. Several applications of such techniques to financial series
modelling and risk management has emerged,
notably~\cite{JDF_Malevergne},
~\cite{JDF_DiasEmbrechts},~\cite{JDF_ChenFanPatton},~\cite{JDF_KoleEtAl},~\cite{JDF_Weiss2011b},
among others.

\medskip

For copula GOF testing, we are only interested in the copula itself,
and the marginal distributions $F_k$, $k=1,\ldots,d$ are seen as
nuisance parameters. Therefore, they are usually replaced by the
marginal empirical cdfs' $F_{n,k}$. Equivalently, the observations
$\Xf_i$, $ i=1,\ldots,n$ are often replaced by their
pseudo-observations $\hat\Uf_{i}:=(F_{n,1}(X_{i,1}),\ldots,
F_{n,d}(X_{i,d}))$. Moreover, for composite zero assumptions, the
chosen estimator $\hat\theta_n$ disturbs the limiting law of the
test statistics most of the time. This difficulty is typical of the
statistics of copulas, and it is a common source of mistakes, as
pointed out in~\cite{JDF_FermScaillet}. For instance,
in~\cite{JDF_BreymannDias}, these problems were not tackled
conveniently and the reported $p$-values are incorrect.
~\cite{JDF_BreymannDias} noticed that the r.v. $\sum_{k=1}^d
[\Phi^{-1}(V_{k})]^2$ follows a $\chi^2(d)$. But it is no more the
case of $\sum_{k=1}^d [\Phi^{-1}(\hat V_{n,k})]^2$, where
$\hat\Vf=\R(\hat\Uf)$. This point has been pointed out
in~\cite{JDF_GenestRem}. A corrected test statistics with reliable
$p$-values has been introduced in~\cite{JDF_DobricSchmid2007}. An
extension of these tests has been introduced
in~\cite{JDF_BergBakken2007}. It implies data-driven weight
functions, to emphasize some regions of underlying the copula
possibly. Its comparative performances are studied
in~\cite{JDF_BergBakken2006} and~\cite{JDF_Berg}.

\medskip

Thus, to the best of our knowledge, all the previous proposed tests procedures have
to rely on bootstrap procedures to evaluate the corresponding
limiting laws under the null. This is clearly a shame, keeping in
mind the simplicity of the law of $\Vf$, after a PIT of the original
dataset (but with {\it known} margins). In practice, we have to work
with (transformed) pseudo-observations $\hat \Vf_{i}$,
$i=1,\ldots,n$. As we said, they are calculated from formulas~(\ref{JDF_PIT}),
replacing unobservable uniformly distributed vectors $\Uf_i$ by pseudo-observations $\hat\Uf_i$, $i=1,\ldots,n$. The vectors $\hat\Vf_i$ are no longer independent and only approximately
uniform on $[0,1]^d$. Nonetheless, test statistics $T_n^{\psi,PIT} =
\psi(\hat\Vf_{1},\ldots,\hat\Vf_{n})$ may be relevant, for convenient real functions
$\psi$. In general and for composite zero assumptions, we are not insured that the law of
$\hat\Vf$, denoted by $C_{\infty,\Vf}$, tends to the independence
copula. If we were able to evaluate $C_{\infty,\Vf}$, a
``brute-force'' approach would still be possible, as in
section~\ref{JDF_sec:2}. For instance and naively, we could
introduce the Kolmogorov-type statistics
$$ T_n^{KM,PIT}:=\sup_{\uf\in (0,1)^d} | \frac{1}{n} \sum_{i=1}^n
\1 (\hat\Vf_{i} \leq \uf) - C_{\infty,\Vf}(\uf) |.$$
 Nonetheless, due to the difficulty to evaluate precisely $C_{\infty,\Vf}$ (by
 Monte-Carlo, in practice), most authors have preferred to reduce the
dimensionality of the problem. By this way, they are able to
tackle more easily the case $d\geq 3$.

\medskip

\subsection{Reductions of dimension}\label{JDF_subsec:5}

Generally speaking, in a GOF test, it is tempting to reduce the
dimensionality of the underlying distributions, for instance from
$d$ to one. Indeed, especially when $d>>1$, the ``brute-force'' procedures based on
empirical processes involve significant analytical or numerical
difficulties in practice. For instance, a Cramer-von-Mises
necessitates the calculation of a $d$-dimensional integral.

\medskip

Formally, a reduction of dimension means replacing the initial GOF
problem ``$\H_0: \text{the copula of $\Xf$ is $C_0$}$'' by
``$\H_0^*: \text{the law of $\psi(\Xf)$ is $G_{\psi,0}$}$'', for
some transformation $\psi:\RR^d \rightarrow \RR^p$, with $p<<d$, and
for some $p$-dimensional cdf $G_{\psi,0}$. As $\H_0$ implies
$\H_0^*$, we decide to reject $\H_0$ when $\H_0^*$ is not satisfied.
Obviously, this reduction of the available information induces a
loss of power, but the practical advantages of this trick often
dominate its drawbacks.

\medskip

For instance, when $p=1$ and if we are able to identify
$G_{\psi,_0}$, it becomes possible to invoke standard univariate GOF
test statistics, or even to use ad-hoc visual procedures like
QQ-plots. Thus, by reducing a multivariate GOF problem to a
univariate problem, we rely on numerically efficient procedures,
even for high dimensional underlying distributions. However, we
still depend on Monte-Carlo methods to evaluate the corresponding
$p$-values. Inspired by~\cite{JDF_SaundersLaud}, we get one of the
most naive method of dimension reduction: replace $T_n^{KS}$ above
by
$$ \tilde T_n^{KS}:=\sum_{\alpha\in (0,1)} | C_n(A_\alpha) -
C_0(A_\alpha)|, \;\;\text{or}\;\; \tilde T_n^{KS}:=\sum_{\alpha\in
(0,1)} | C_n(\hat A_\alpha) - C_{\hat\theta_n}(\hat A_\alpha)|,$$ where
$(A_\alpha)_{\alpha\in (0,1)}$ is an increasing sequence of subsets
in $[0,1]^d$ s.t. $A_\alpha=\{ \uf \in [0,1]^d | C_0(\uf)\leq \alpha
\}$ and $\hat A_\alpha=\{ \uf \in [0,1]^d | C_{\hat\theta_n}(\uf)\leq \alpha
\}$.

\medskip

To revisit a previous example and with the same notations,~\cite{JDF_DobricSchmid2007} considered particular test
statistics $T_n^{\psi, PIT}$ based on the variables $ \hat Z_i :=
\sum_{k=1}^d \Phi(\hat V_{i,k})^{-1}$, $i=1,\ldots,n$. If the
margins $F_k$, $k=1,\ldots,d$, and the true copula $C_0$ were known,
then we were be able to calculate $Z_i:=\sum_{k=1}^d
\Phi(V_{i,k})^{-1}$ that tends in law towards a chi-square law of dimension $d$ under the null. Since it is not the
case in practice, the limiting law of $\hat Z_i$ is unknown, and it
has to be evaluated numerically by simulations. It is denoted by
$F_{\hat Z}$. Therefore,~\cite{JDF_DobricSchmid2007}
propose to test
$$\H_0^*: \text{the asymptotic law of $T_n^{\psi,PIT}$ is a given cdf $F_{\psi}$ (to be estimated)},$$
where $T_n^{\psi,PIT}$ is defined by usual (univariate)
Kolmogorov-Smirnov, Anderson-Darling or Cramer-von-Mises test
statistics. For instance,
$$ T_n^{AD,PIT}:= n\int \frac{(F_{n,\hat Z} - F_{0,\hat Z})^2}{F_{0,\hat Z}(1- F_{0,\hat
Z})},$$ where $F_{n,\hat Z}$ is the empirical cdf of the pseudo
sample $\hat Z_1,\ldots,\hat Z_n$. Note that $F_{n,\hat Z}$ and
$F_{0,\hat Z}$ depend strongly on the underlying cdf of $\Xf$, its
true copula $C_0$, the way marginal cdfs' have been estimated to
get pseudo-observations (empirical or parametric estimates) and
possibly the particular estimate $\hat\theta_n$.

\medskip

Beside the PIT idea, there exist a lot of possibilities of dimension
reductions potentially. They will provide more or less relevant test
statistics, depending on the particular underlying parametric family
and on the empirical features of the data. For instance, in the
bivariate case, Kendall's tau $\tau_K$ or Spearman's rho $\rho_S$
may appear as nice ``average'' measures of dependence. They are just
single numbers, instead of a true $2$-dimensional function like $C_
n$. Therefore, such a GOF test may be simply
$$\H_0^*: \hat\tau_K = \tau_{K,C_0},$$
where $\tau_{K,C_0}=4 \EE_{C_0}[C_0(\Uf)] - 1$ is the Kendall's
tau of the true copula $C_0$, and $\hat\tau_{K}$ is an estimate of
this measure of dependence, for instance its empirical counterpart
$$ \hat\tau_{K,n} := \frac{2}{n(n-1)}[\text{number of concordant pairs of observations} - \text{number of discordant pairs}].$$
Here, we can set $T_n^{KTau}:=n(\hat \tau_{K,n} - \tau_{C_0})^2$, or
$T_n^{KTau}:=n(\hat \tau_{K,n} - \tau_{C_{\hat\theta_n}})^2$ in the
case of composite assumption. Clearly, the performances of all these
tests in terms of power will be very different and there is no hope
to get a clear hierarchy between all of them. Sometimes, it will be
relevant to discriminate between several distributions depending on
the behaviors in the tails. Thus, some adapted summaries of the
information provided by the underlying copula $C$ are required, like
tail-indices for instance (see~\cite{JDF_Nel99} e.g.). But in every
case, their main weakness is a lack of convergence against a large
family of alternatives. For instance, the previous test $T_n^{KTau}$
will not be able to discriminate between all copulas that have the
same Kendall's tau $\tau_{K,C_0}$. In other words, this dimension
reduction is probably too strong, most of the time: we reduce a
$d$-dimensional problem to a real number. It is more fruitful to
keep the idea of generating a univariate process, i.e., going from a
dimension $d$ to a dimension one. This is the idea of Kendall's
process (see below).

\medskip

Another closely related family of tests is based on the comparison
between several parameter estimates. They have been called
``moment-based'' GOF test statistics
(see~\cite{JDF_Shih},~\cite{JDF_GenestQuessyRem2006b},~\cite{JDF_BergQuessy}).
In their simplest form, assume a univariate unknown copula parameter
$\theta$, and two estimation equations (``moments'') such that
$m_1=r_1(\theta)$ and $m_2=r_2(\theta)$ (one-to-one mappings). Given
empirical counterparts $\hat{m}_k$ of $m_k$,
$k=1,2$,~\cite{JDF_Shih} has proposed the copula GOF test
$$ T_n^{moment}:= \sqrt{n}\left\{r_1^{-1}(\hat m_1)-r_2^{-1}(\hat m_2)
\right\}.$$ Typically, some estimating equations are provided by
Kendall's tau and Spearman's rho, that have well-known empirical
counterparts. Nonetheless, other estimates have been proposed, as
the pseudo-maximum likelihood (also called ``canonical maximum
likelihood''). To deal with multi-dimensional parameters $\theta$,
estimating equations can be obtained by the equality between the
hessian matrix and minus the expected outer product of the score
function. This is the idea of White's specification test
(see~\cite{JDF_White}), adapted to copulas in~\cite{JDF_Prokhorov}.

\medskip

\subsection{Kendall's process}\label{JDF_subsec:6}

This is another and well-known example of dimension reduction
related to copula problems. Let $C$ be the copula of an arbitrary
random vector $\Xf\in \RR^d$. Define the univariate cdf
$$ K(t):= \PP( C(\Uf)\leq t),\;\; \forall t\in \RR,$$
where, as usual, we set $\Uf=(F_1(X_1),\ldots,F_d(X_d))$. The
function $K$ depends on $C$ only. Therefore, this univariate
function is a ``summary'' of the underlying dependence structure
given by $C$. It is called the Kendall's dependence function of
$C$. An empirical counterpart of $K$ is the empirical Kendall's
function
$$ K_n(t):=\frac{1}{n }\sum_{i=1}^n \1 (C_n(\hat\Uf_i)\leq t),$$
with pseudo-observations $\hat\Uf_1,\ldots,\hat\Uf_n$. The
associated Kendall's process is simply given by $\KK_n=\sqrt{n}(K_n
- K)$, or $\hat\KK_n=\sqrt{n}(K_n - K(\hat\theta_n,\cdot))$ when the
true copula is unknown but belongs to a given parametric family. The
properties of Kendall's processes has been studied in depth
in~\cite{JDF_BarbeEtAl},~\cite{JDF_GhoudiRem1998},
and~\cite{JDF_GenestQuessyRem2006a} particularly. In the latter
papers, the weak convergence of $\KK_n$ towards a continuous
centered Gaussian process in the Skorohod space of cadlag functions
is proved, for convenient consistent sequences of estimates
$\hat\theta_n$. Its variance-covariance function is complex and
copula dependent. It depends on the derivatives of $K$ w.r.t. the
parameter $\theta$ and the limiting law of $\sqrt{n}(\hat\theta_n -
\theta_0)$.

\medskip

Then, there are a lot of possibilities of GOF tests based on the
univariate function $K_n$ or the associated process $\KK_n$. For
instance,~\cite{JDF_WangWells} introduced a test statistics based on
the $L^2$ norm of $\KK_n$. To be specific, they restrict themselves
to bivariate Archimedean copulas, but allow censoring. That is why
their GOF test statistics $T_n^{L2,Kendall}=\int_\xi^1 | \KK_n |^2$
involves an arbitrary cut-off point $\xi>0$. Nonetheless, the idea
of such a statistics is still valid for arbitrary dimensions and
copulas. It has been extended in~\cite{JDF_GenestQuessyRem2006a},
that considers
$$T_n^{L2,Kendall}:=\int_0^1 | \KK_n(t) |^2
k(\hat\theta_n,t)\,dt,\;\; \text{and}\;\;
T_n^{KS,Kendall}:=\sup_{t\in [0,1]} | \KK_n(t) |,
$$ where $k(\theta,\cdot) $ denotes the density of $C(\Uf)$ w.r.t.
to the Lebesgue measure (i.e. the derivative of $K$), and
$\hat\theta_n$ is a consistent estimate of the true parameter
under the null.

\medskip

Nonetheless, working with $\KK_n$ or $\hat\KK_n$ instead of $\CC_n$
or $\hat\CC_n$ respectively is not the panacea. As we said, the
dimension reduction is not free of charge, and testing $\H_0^*$
instead of $\H_0$ reduces the ability to discriminate between copula
alternatives. For instance, consider two extreme-value copulas
$C_{1}$ and $C_2$, i.e., in the bivariate case,
$$ C_j(u,v)= \exp\left(\ln(uv)A_j(\frac{\ln u}{\ln uv}) \right),\,\, j=1,2,$$
for some Pickands functions $A_1$ and $A_2$ (convex functions on
$[0,1]$, such that $\max(t,(1-t))\leq A_j(t) \leq 1$ for all $t\in
[0,1]$). As noticed in~\cite{JDF_GhoudiRem1998}, the associated
Kendall's functions are
$$ K_j(t)= t - (1-\tau_{K,j})t\ln t,\; t\in (0,1),$$
where $\tau_{K,j}$ denotes the Kendall's tau of $C_j$. Then, if the
two Kendall's tau are the same, the corresponding Kendall's
functions $K_1$ and $K_2$ are identical. Thus, a test of $\H_0^*:
K=K_0$ will appear worthless if the underlying copulas are of the
extreme-value type.

\medskip

In practice, the evaluation of the true Kendall function $K_0$ under
the null may become tedious, or even unfeasible for a lot of copula
families. Therefore, ~\cite{JDF_BergBakken2006} proposed to apply
the previous Kendall process methodology to random vectors obtained
through a PIT in a preliminary stage, to ``stabilize'' the limiting
law under the null. In this case, $K_0$ is always the same: the
Kendall function associated to the independence copula $C_{\perp}$.
This idea has been implemented in~\cite{JDF_BeauGenestRem}, under
the form of Cramer-von-Mises GOF test statistics of the type
$$ T_n^{CvM,PIT}:=n\int \left(D_n(\uf) - C_\perp(\uf)\right)^2\,
dD_n(\uf)=\sum_{i=1}^n \left(D_n(\hat\Uf_i)-C_\perp (\hat\Uf_i)
\right)^2,$$ were $D_n(\uf)=n^{-1}\sum_{i=1}^n \1(\hat\Uf_i\leq
\uf)$ is the empirical cdf associated to the pseudo-observations
of the sample. Nonetheless, the limiting behavior of all these
test statistics are not distribution-free for composite zero
assumptions, and limiting laws have to be evaluated numerically by
Monte-Carlo methods (as usual).

\medskip

Note that~\cite{JDF_Quessy} have proposed a similar idea, but
based on Spearman's dependence function $L$ instead of Kendall's
dependence function. Formally, $L$ is defined by
$$ L(u):= \PP \left(C_\perp (\Uf)\leq u   \right)=\PP \left(\prod_{k=1}^d F_k(X_k)\leq u   \right),\;\; \forall u\in
[0,1].$$ When working with a random sample, the empirical
counterpart of $L$ is then
$$ \hat L_n(u):= \frac{1}{n} \sum_{i=1}^n \1 \left(C_\perp (\hat\Uf_i)\leq u
\right),$$ and all the previous GOF test statistics may be
applied. For instance,~\cite{JDF_Berg} proposed to use the
Cramer-von-Mises statistic
$$ T_n^{L,CvM}:= \int_0^1 \left(\hat L_n -  L_{\hat\theta_n}
\right)^2\, \hat L_n(du),$$ where $L(\theta)$ is the Spearman's
dependence function of an assumed copula $C_\theta$, and
$\hat\theta_n$ is an estimate of the true parameter under the zero
assumption.

\medskip

\section{GOF tests for some particular classes of copulas}\label{JDF_sec:4}

Beside omnibus GOF tests, there exist other test statistics that are
related to particular families of copulas only. We will not study
such GOF tests when they are related to particular finite-dimensional
parametric families (to decide whether $C_0$ is a Gaussian copula, for
instance). Nonetheless, in this section, we will be interested in a
rather unusual GOF problem: to say whether $C_0$ belongs to a
particular infinite-dimensional parametric family of copulas. Among
such large families, some of them are important in practice: the
Archimedean family, the elliptical one, extreme-value copulas, vines, hierarchical
Archimedean copulas etc.

\medskip

\subsection{Testing the Archimedeanity}\label{JDF_subsec:7}

All the previously proposed test statistics can be applied when $\C$
is an assumed particular Archimedean family, as
in~\cite{JDF_WangWells},~\cite{JDF_SavuTrede}... Other test
statistics, that are based on some analytical properties of
Archimedean copulas, have been proposed too
(~\cite{JDF_HeringHofert}, for instance).
Interestingly,~\cite{JDF_GenestRivest} proposed a graphical
procedure for selecting a Archimedean copula (among several
competitors), through a visual comparison between the empirical
Kendall's function $K_n$ and an estimated Kendall function obtained
under a composite null hypothesis $\H_0$.

\medskip

Now, we would like to test ``$ \H_0: C \;\text{is Archimedean}$''
against the opposite, i.e., without any assumption concerning a
particular parametric family. This problem has not received a lot of
attention in the literature, despite its practical importance.

\medskip

Consider first the (unknown) generator $\phi$ of the underlying
bivariate copula $C$, i.e. $C(\uf)=\phi^{-1}(\phi(u_1)+\phi(u_2))$
for every $\uf=(u_1,u_2)\in [0,1]^2$. ~\cite{JDF_GenestRivest}
proved that $V_1:=\phi(F_1(X_1))/\{\phi(F_1(X_1))+\phi(F_2(X_2))\}$
is uniformly distributed on $(0,1)$ and that
$V_2:=C(F_1(X_1)),F_2(X_2))$ is distributed as the Kendall's
dependence function $K(t)=t-\phi(t)/\phi'(t)$. Moreover, $V_1$ and
$V_2$ are independent. Since $K$ can be estimated empirically, these
properties provide a way of estimating $\phi$ itself (by $\phi_n$).
Therefore, as noticed in the conclusion of~\cite{JDF_GenestRivest},
if the underlying copula is Archimedean, then the r.v.
$$ \hat
V_1:=\phi_n(F_{1,n}(X_1))/\{\phi(F_{1,n}(X_1))+\phi(F_{2,n}(X_2))\}$$
should be distributed uniformly on $(0,1)$ asymptotically. This
observation can lead to some obvious GOF test procedures.

\medskip

Another testing strategy starts from the following property,
proved in~\cite{JDF_Nel99}: a bivariate copula $C$ is Archimedean
iff it is associative (i.e. $C(u_1,C(u_2,u_3))=C(C(u_1,u_2),u_3)$
for every triplet $(u_1,u_2,u_3)$ in $[0,1]^3$) and satisfies the
inequality $C(u,u)<u$ for all $u\in (0,1)$.
This property, known as Ling's Theorem (see~\cite{JDF_Ling}), has been extended in an arbitrary dimension $d>2$ by~\cite{JDF_Stupnanova}.
Then,~\cite{JDF_Jaworski} proposed to test the associativity of
$C$ to check the validity of the Archimedean zero assumption. For
every couple $(u_1,u_2)$ in $(0,1)^2$, he defined the test
statistics
$$
\T_n^J(u_1,u_2):=\sqrt{n}\left\{C_n(u_1,C_n(u_2,u_2))-C_n(C_n(u_1,u_2),u_2)\right\}.$$
Despite its simplicity, the latter pointwise approach is not
consistent against a large class of alternatives. For instance,
there exist copulas that are associative but not Archimedean.
Therefore,~\cite{JDF_BucherDette2012} revisited this idea, by
invoking fully the previous characterization of Archimedean copulas.
To deal with associativity, they introduced the trivariate process
$$ \T_n(u_1,u_2,u_3):=\sqrt{n}\left\{C_n(u_1,C_n(u_2,u_3))-C_n(C_n(u_1,u_2),u_3)\right\},$$
and proved its weak convergence in $\ell^\infty([0,1]^3)$.
Cramer-von-Mises $T_n^{CvM}$ and Kolmogorov-Smirnov $T_n^{KS}$
test statistics can be build on $\T_n$. To reject associative
copulas that are not Archimedean, these statistics are slightly
modified to get
$$ \tilde T_n^{CvM}:= T_n^{CvM}+n^\alpha \psi \left(\max\left\{\frac{i}{n}(1-\frac{i}{n}): C_n(\frac{i}{n},\frac{i}{n})=\frac{i}{n}
 \right\}   \right),$$
for some chosen constant $\alpha\in (0,1/2)$ and some increasing
function $\psi$, $\psi(0)=0$. Therefore, such final tests are
consistent against all departures from Archimedeanity.

\medskip

Unfortunately, the two previous procedures are limited to bivariate
copulas, and their generalization to higher dimensions $d$ seems to
be problematic.

\medskip

\subsection{Extreme-value dependence}\label{JDF_subsec:8}

As we have seen previously, bivariate extreme-value copulas are written as
\begin{equation}
 C(u,v)=\exp\left\{ \ln(uv) A( \frac{\ln (v)}{\ln (uv)} )
\right\},
\label{extremevalue}
\end{equation}
 for every $u,v$ in $(0,1)$, where $A:[0,1]\rightarrow
[1/2,1]$ is convex and satisfies $ \max(t,1-t) \leq A(t) \leq 1$ for
every $t\in [0,1]$. Therefore, such copulas are fully parameterized
by the so-called Pickands dependence function $A$, that is
univariate. Extreme-value copulas are important in a lot of fields
because they characterize the large-sample limits of copulas of
componentwise maxima of strongly mixing stationary sequences
(\cite{JDF_Deheuvels1984},~\cite{JDF_Hsing}, and the recent survey~\cite{JDF_Gudendorf}).
Then, it should be of interest to test whether whether the underlying copula can be represented
by~(\ref{extremevalue}), for some unspecified dependence function
$A$.

 \medskip

Studying the Kendall's process associated to an extreme-value copula
$C$,~\cite{JDF_Ghoudi1998} have noticed that, by setting
$W:=C(U_1,U_2)$, we have $ K(t) = P(W\leq t) = t - (1-\tau) t \ln
(t),$ for every $t\in (0,1)$, where $\tau$ is the underlying
Kendall's tau. Moreover, they show that the moments of $W$ are $
E[W^i]=(i\tau +1)/(i+1)^2$, for all $i\geq 1$. Therefore, under
$\H_0$, $-1 + 8E[W] -9E[W^2]=0$. Then they proposed a test
(that the underlying copula is extreme-value) based on an empirical
counterpart of the latter relation: set
$$ T_n := -1 + \frac{8}{n(n-1)}\sum_{i\neq j}
I_{ij}-\frac{9}{n(n-1)(n-2)}\sum_{i\neq j\neq k}I_{ij}I_{kj},$$
where $I_{ij}:= \1(X_{i,1}\leq X_{j,1},X_{i,2}\leq X_{j,2} )$, for
all $i,j\in \{1,\ldots,n\}$. Under $\H_0$, the latter test statistic
is asymptotically normal. Its asymptotic variance has
been evaluated in~\cite{JDF_BenGhorbal}.~\cite{JDF_Quessy2011} has
provided extensions of this idea towards more higher order moments
of $W$. 

\medskip

These approaches rely on the so-called "reduction of dimension"
techniques (see Section...). To improve the power of GOF tests, it
would be necessary to work in functional spaces, i.e. concentrate on
empirical counterparts of extreme-value copulas, or, equivalently,
of the functions $A$ themselves. 
For instance,~\cite{JDF_Quessy2011} proposed a Cramer-von-Mises GOF test, based on
the Kendall's function $K$ above.
More generally, several estimates of the
Pickands dependence function are available, but most of them rely on
the estimation of marginal distributions: see section 9.3
in~\cite{JDF_Beirlant} or~\cite{JDF_Abdous}. Nonetheless,~\cite{JDF_GenestKojadinovic2011} have built "pure" copula GOF test
statistics, i.e. independent from margins, by invoking empirical
counterparts of the Pickands function introduced
in~\cite{JDF_GenestSegers2009}: given our previous notations,
\begin{enumerate}
\item define the pseudo-observations
$$\tilde U_{i}:= nF_{n,1}(X_{i,1})/(n+1), \; \tilde V_{i}:=
nF_{n,1}(X_{i,2})/(n+1);$$
\item define the r.v. $\hat S_i :=-\ln \tilde U_i$ and $\hat T_i :=-\ln \tilde
V_i$;
\item for every $i=1,\ldots,n$, set $\hat\xi(0):=\hat S_i$, and $\hat\xi(1):=\hat T_i$. Moreover, for every $t\in (0,1)$, set
$$\hat \xi_i(t) := \min\left( \frac{\hat S_i}{1-t},\frac{\hat T_i}{t}   \right).$$
\item Two estimates of $A$ are given by
$$  A^P_n(t) :=\left[ n^{-1}\sum_{i=1}^n \hat \xi_i(t) \right]^{-1}\; \text{and}\;\; A_n^{CFG}(t):=\exp\left(-\gamma - n^{-1}\sum_{i}^n  \ln \hat\xi_i(t)    \right),$$
where $\gamma $ denotes the Euler constant.
\end{enumerate}
The two latter estimates are the "rank-based" version of those
proposed in~\cite{JDF_Pickands} and~\cite{JDF_Caperaa} respectively.

\medskip

There is an explicit one-to-one mapping between $A^P_n$ (resp.
$A_n^{CFG}$) and the empirical copula $C_n$. Therefore, after
endpoint corrections,~\cite{JDF_GenestSegers2009} have exhibited the
weak limit of the corresponding processes $\Abb_n^P:=\sqrt{n}(A_n^P -
A)$ and $\Abb_n^{CFG}:=\sqrt{n}(A_n^{CFG} - A)$. Working with the two
latter processes instead of $\CC_n$, a lot of GOF tests can be
built. For instance~,~\cite{JDF_GenestKojadinovic2011} have detailed an
Anderson-Darling type test based on the $L^2$ norm of $\Abb_n^P$ and
$\Abb_n^{CFG}$, even under composite null assumptions. 

\medskip
    
In the same
vein, another strategy has been proposed
in~\cite{JDF_KojadinovicSegersYan}: there is an equivalence between
extreme-value copula $C$ and max-stable copulas, i.e. copulas for
which $C(\uf)^r=C(\uf^r)$, for every $\uf\in [0,1]^d$ and $r\in
\RR^+$. By setting $\DD_n(\uf):=\sqrt{n}(\{C_n(\uf^{1/r})\}^r -
C_n(\uf))$, for all $\uf\in[0,1]^d$ and every
$r>0$,~\cite{JDF_KojadinovicSegersYan} have built some tests based
on the limiting law of the joint process
$(\DD_{n,r_1},\ldots,\DD_{n,r_p})$ for an arbitrary integer $p$.

\medskip

\subsection{Pair-copula constructions}
\label{JDF_subsec:9}

In the recent years, a lot of effort has been devoted to the
construction of $d$-dimensional copulas,  $d>2$, as combinations of
several $2$-dimensional copulas. Some authors have enriched the
Archimedean copula class: Hierarchical, nested or multiplicative
Archimedean copulas. Among others,
see~\cite{JDF_Joe},~\cite{JDF_Whelan},~\cite{JDF_Morillas},~\cite{JDF_SavuTrede2006},~\cite{JDF_Okhrin}.
Other authors have studied the large class of vines: D-vines,
C-vines, regular vines more generally
(see~\cite{JDF_Aas},~\cite{JDF_Czado}, e.g.). Inference, simulation
and specification techniques have made significant progress to deal
with these families of models $\F$. This advances provide large
classes of very flexible copulas.

\medskip

We will not discuss in depth the way of choosing the best
Hierarchical Archimedean copula or the best D-vine, for a given
data. Apparently, every proposition in this stream of the literature
follows the same steps:
\begin{enumerate}
\item[(i)] Assume an underlying class of models $\F$ (D-vine, for instance);
\item[(ii)] Choose the potential bivariate families of copulas that may
appear in the construction;
\item[(iii)] Evaluate the best structure (a  network, or a tree), and estimate the associated
bivariate copulas (simultaneously, in general).
\end{enumerate}
Mathematically, we can nest this methodology inside the previous
general GOF copula framework detailed above. Indeed, the copula
candidates belong to a finite dimensional parametric family, even if
the dimension of the unknown parameter $\theta$ can be very large.
Obviously, authors have developed ad-hoc procedures to avoid such a
violent approach of GOF testing: see~\cite{JDF_CzadoEtAl}
or~\cite{JDF_Dissman} for vine selection, for instance.

\medskip

At the opposite, there is no test of the slightly different and more
difficult GOF problem
$$ \H_0: C \;\text{belongs to a given class}\; \F.$$
For instance, a natural question would be to test whether an
underlying copula belongs to the large (and infinite dimensional!)
class of Hierarchical Archimedean copulas. To the best of our
knowledge, this way of testing is still a fully open problem.

\medskip

\section{GOF copula tests for multivariate time series}\label{JDF_sec:5}

One limiting  feature of copulas is the difficulty to use them in
the presence of multivariate dependent vectors $(\Xf_n)_{n\in
\ZZ}$, with $\Xf_n\in \RR^d$. In general, the ``modeler problem''
is to specify the full law of this process, i.e., the joint laws
$(\Xf_{n_1},\ldots,\Xf_{n_p})$ for every $p$ and every indices
$n_1,\ldots,n_p$ and in a consistent way. Applying the copula
ideas to such a problem seems to be rather natural
(see~\cite{JDF_Patton2009} for a survey). Nonetheless, even if we
restrict ourselves to stationary processes, the latter task is far
from easy.

\medskip

A first idea is to describe the law of the vectors
$(\Xf_{m},\Xf_{m+1},\ldots,\Xf_n)$ with copulas directly, for every
couple $(m,n)$, $m<n$. This can be done by modeling separately (but
consistently) $d(n-m+1)$ unconditional margins plus a
$d(n-m+1)$-dimensional copula. This approach seems particularly
useful when the underlying process is stationary and Markov (see~\cite{JDF_ChenFan2006a} for the general procedure). But the
conditions of Markov coherence are complex
(see~\cite{JDF_Ibragimov}), and there is no general GOF strategy in
this framework, to the best of our knowledge.

\medskip

A more usual procedure in econometrics is to specify a multivariate
time-series model, typically a linear regression, and to estimate
residuals, assumed serially independent:
see~\cite{JDF_ChenFan2006b}, that deals with a GARCH-like model with
diagonal innovation matrix. They showed that estimating the copula
parameters using rank-based pseudo-likelihood methods with the ranks
of the residuals instead of the (non-observable) ranks of
innovations, leads to the same asymptotic distribution. In
particular, the limiting law of the estimated copula parameters does
not depend on the unknown parameters used to estimate the
conditional means and the conditional variances. This is very useful
to develop goodness-of-fit tests for the copula family of the
innovations.~\cite{JDF_Remillard} extended these results: under
similar technical assumptions, the empirical copula process has the
same limiting distribution as if one would have started with the
innovations instead of the residuals. As a consequence, a lot of
tools developed for the serially independent case remain valid for
the residuals. However, that is not true if the stochastic
volatility is genuinely non-diagonal.

\medskip

A third approach would be to use information on the marginal
processes themselves. This requires to specify conditional marginal
distributions, instead of unconditional margins as above in the
first idea. This would induce a richer application of the two-step
basic copula idea i.e., use ``standard'' univariate processes as
inputs of more complicated multivariate models:
\begin{enumerate}
\item for every $j=1,\ldots,d$, specify the law of $X_{n,j}$ knowing
the past values $X_{n-1,j},X_{n-2,j}, \ldots$;
\item specify (and/or estimate) relevant
dependence structures, ``knowing'' these univariate underlying
processes, to recover the entire process $(\Xf_n)_{n\in \ZZ}$.
\end{enumerate}

Using similar motivations, Patton
(\cite{JDF_Patton2001},~\cite{JDF_Patton2006}) introduced so-called
conditional copulas, which are associated with conditional laws in a
particular way. Specifically, let $\Xf=(X_1,\ldots,X_d)$ be a random
vector from $(\Omega,\A_0,\PP)$ to $\RR^d$. Consider some arbitrary
sub-$\sigma$-algebra $\A\subset \A_0$. A conditional copula
associated to $(\Xf,\A)$ is a $\B([0,1]^d)\otimes \A$ measurable
function $C$ such that, for any $x_1, \ldots, x_d \in \mathbb{R}$,
$$ \PP\left( \Xf\leq  \xf |\A \right) = C\left\{ \PP(X_1\leq
x_1|\A),\ldots, \PP(X_d\leq x_d|\A) |\A \right\}.$$ The random
function $C( \cdot |\A)$ is uniquely defined on the product of the
values taken by $x_j\mapsto \PP ( X_j\leq x_j\ |\ \A)(\omega)$,
$j=1,\ldots,d$, for every realization $\omega\in \A$. As in the
proof of Sklar's theorem, $C( \cdot |A)$ can be extended on
$[0,1]^d$ as a copula, for every conditioning subset of events
$A\subset\A$.

\medskip

In Patton's approach, it is necessary to know/model each margin,
knowing all the past information, and not only the past observations
of each particular margin. Nonetheless, practitioners often have
good estimates of the conditional distribution  of each  margin,
conditionally given its own past, i.e.,  $\PP (X_{n,j}\leq x_j |
\A_{n,j} )$, $j=1,\ldots,d$, by setting $\A_{n,j} = \sigma(
X_{n-1,j},X_{n-2,j},\ldots)$. To link these quantities with the
(joint) law of $\Xf_n$ knowing its own past, it is tempting to write
$$\PP\left( \Xf_n\leq  \xf |\A_n \right) = C^\ast\left\{ \PP (
X_{1,n}\leq x_1|\A_{n,1}),\ldots, \PP(X_{d,n}\leq x_d|\A_{n,d}
)\right\},$$ for some random function $C^\ast:[0,1]^d\longrightarrow
[0,1]$ whose measurability would depend on $\A_n$ and on the
$\A_{n,j}$, $j=1,\ldots,d$. Actually, the latter function is a
copula only if the process $(X_{k,n},k\neq j)_{n\in \ZZ}$ does not
``Granger-cause'' the process $(X_{j,n})_{n\in \ZZ}$, for every
$j=1,\ldots,d$. This assumption that each variable depends on its
own lags, but not on the lags of any other variable, is clearly
strong, even though it can be accepted empirically; see the
discussion in~\cite{JDF_Patton2009}, pp.\ 772--773.
Thus,~\cite{JDF_FermWeg} has extended Patton's conditional copula
concept, by defining so-called pseudo-copulas, that are simply cdf
on $[0,1]^d$ with arbitrary margins. They prove:

\begin{theorem}
For any sub-algebras $\B,\A_1,\ldots,\A_d$ such that $\A_j \subset
\B$, $j=1,\ldots,d$, there exists a random function
$C:[0,1]^d\times \Omega \longrightarrow [0,1]$ such that
\begin{eqnarray*}
\PP( \Xf \leq \xf\ |\ \B)(\omega) &=& C\left\{ \PP (X_{1}\leq x_1\
|\ \A_1)(\omega),\ldots,
\PP( X_{d}\leq x_d\ |\ \A_d)(\omega)\, ,\, \omega\right\} \nonumber \\
&\equiv& C\left\{\PP(X_{1}\leq x_1\ |\ \A_1),\ldots, \PP(X_{d}\leq
x_d\ |\ \A_d)\right\}(\omega), \label{defcopula}
 \end{eqnarray*}
for every $\xf=(x_1,\ldots,x_d)\in\RR^d$ and almost every
$\omega\in\Omega$. This function $C$ is $\B([0,1]^d)\otimes \B$
measurable. For almost every $\omega\in\Omega$, $C(\cdot,\omega)$
is a pseudo-copula and is uniquely defined on the product of the
values taken by $x_j\mapsto \PP( X_j\leq x_j\ |\ \A_j)(\omega)$,
$j=1,\ldots,d$.
\end{theorem}

If $C$ is unique, it is called the conditional
$(\A,\B)$-pseudo-copula associated with $\X$ and denoted by
$C(\cdot|\A,\B)$. Actually, $C(\cdot|\ \A,\B)$ is a copula iff
\begin{equation}
\PP( X_j\leq x_j\ |\ \B)=\PP( X_j\leq x_j \ |\ \A_j)\quad {\rm
a.e.} \label{truecop}
\end{equation}
for all $j=1,\ldots,d$ and $\xf\in\RR^d$. This means that $\B$
cannot provide more information about $X_j$ than $\A_j$, for every
$j$. Patton's conditional copula corresponds to the particular
case $\B=\A_1=\cdots=\A_d$, for which~(\ref{truecop}) is clearly
satisfied.

\medskip

One key issue is to state if pseudo-copulas depend really on the
past values of the underlying process, i.e., to test their
constancy, an assumption often made in practice.
In~\cite{JDF_FermWeg}, they estimate nonparametrically
conditional pseudo-copulas, including Patton's conditional copulas
as a special case, and test their constancy with respect to their
conditioning subsets. Here, we specify their technique.

\medskip

For a stationary and strongly mixing process $(\Xf_n)_{n\in\ZZ}$, we
restrict ourselves to conditional sub-algebras $\A_n$ and $\B_n$
that are defined by a finite number of past values of the process,
typically $(\Xf_{n-1},\Xf_{n-2},\ldots,\Xf_{n-p})$ for some $p\geq
1$. The dependence of $\A$ and $\B$ with respect to past
values $\yf$ will be implicit hereafter.
Formally,~\cite{JDF_FermWeg} consider the test of several null
hypothesis:
\begin{enumerate}
\item[(a)]
$$  \H_0^{(1)} : \mbox{ For every } \yf,\;C(\cdot\ |\ \A,\B)=C_0(\cdot),$$
against
\[ \H_a : \mbox{ For some } \yf, \; C(\cdot\ |\A,\B)\neq C_0(\cdot),
\]
where $C_0$ denotes a fixed pseudo-copula function. In this
case, $ \H_0^{(1)}$ means that the underlying conditional
$(\A,\B)$-pseudo-copula is in fact a true copula, independent of
the past values of the process.
\item[(b)]
 \begin{multline*}
\H_0^{(2)} : \mbox{ There exists a parameter } \theta_0 \mbox{
such that } C(\cdot | \A,\B)=C_{\theta_0}\in \C, \, \mbox{ for
every } \yf,  \hspace{2cm}
\end{multline*}
where $\C=\{C_\theta,\theta\in \Theta\}$ denotes some parametric
family of pseudo-copulas.
\item[(c)]
 \begin{eqnarray*}
\lefteqn{  \H_0^{(3)} : \mbox{ For some function } \theta
  (\yf)=\theta(\A,\B), \mbox{ we have } }\\
  & & C(\cdot |
  \A,\B)=C_{\theta(\yf)}\in \C,\, \mbox{ for every } \yf.
 \end{eqnarray*}
\end{enumerate}

The latter assumption says that the conditional pseudo-copulas stay
inside the same pre-specified parametric family of pseudo-copulas
(possibly copulas), for different observed values in the past.~\cite{JDF_FermWeg} proposed a fully nonparametric
estimator of the conditional pseudo-copulas, and derived its limiting distribution. This provides a framework for
``brute-force'' GOF tests of multivariate dynamic dependence
structures (conditional copulas, or even pseudo-copulas), similarly
to what has been done in section~\ref{JDF_sec:2}.

\medskip

~\cite{JDF_FermWeg} stated the equivalent of the empirical processes $\CC_n$ or
$\hat\CC_n$. Use the short-hand notation
$\Xf_m^n$ for the vector $(\Xf_m,\Xf_{m+1},\ldots,\Xf_{n})$.
Similarly, write $\Xf_{m,j}^n=(X_{m,j},\ldots,X_{n,j})$. Assume that
every conditioning set $\A_{n,j}$ (resp. $\B_n$) is related to the
vector $\Xf_{n-p,j}^{n-1}$ (resp. $\Xf_{n-p}^{n-1}$). Specifically,
consider the events $(\Xf_{n-p}^{n-1}=\yf^\ast)\in \B_n $, with
$\yf^\ast=(\yf_1,\ldots,\yf_p)$, and
$(\Xf_{n-p,j}^{n-1}=\yf_j^\ast)\in \A_{n,j}$, with $\yf_j^\ast =
(y_{1j},\ldots,y_{pj})$.
Their nonparametric estimator of the  pseudo-copula is based on a
standard plug-in technique that requires estimates of the joint
conditional distribution
\[  m({\bf{x}}\ |\ \yf^\ast)= \PP\left(  {\bf{X}}_p \le {\bf{x} } \ |\ {\bf{X}}_0^{p-1}=\yf^\ast \right),
\]
and of conditional marginal cdf's
\[ m_j(x_j\ |\ \yf_j^\ast)= \PP\left( X_{pj} \le x_j \ |\ {\bf{X}}_{0,j}^{p-1}=\yf_j^\ast \right),\quad j=1,\ldots,d.
\]

Let $F_{nj}$ be the (marginal) empirical distribution function of
$X_j$, based on the $(X_{1,j},\ldots,X_{n,j})$. For convenient
kernels $K$ and $\bar K$, set
 $$
K_h({\bf x})=h^{-pd}
K\left(\frac{x_1}{h},\cdots,\frac{x_{pd}}{h}\right),\;\;
\text{and}\;\; \quad \bar K_{\bar h}({\bf x})=\bar h^{-p} \bar
K\left(\frac{x_1}{\bar h},\cdots,\frac{x_{p}}{\bar h}\right).
 $$
For every $\xf\in \RR^d$ and $\yf^\ast\in \RR^{pd}$, estimate the
conditional distribution $m({\bf{x}}\ |\ \yf^\ast)= \PP\left(
{\bf{X}}_p \le {\bf{x} } \ |\ {\bf{X}}_0^{p-1}=\yf^\ast \right)$
by
\[ m_{n}(\xf\ |\ \yf^\ast)= \frac{1}{n-p} \sum_{\ell=0}^{n-p} K_n ( \Xf_{\ell}^{\ell+p-1} )
\1 ( \Xf_{\ell+p} \leq \xf),\] where
\begin{multline*}
 K_n ( \Xf_{\ell}^{\ell+p-1} ) = K_h \{  F_{n1}(X_{\ell 1})-
F_{n1}(y_{11}),\ldots, F_{nd}(X_{\ell d})- F_{nd}(y_{1d}),\ldots, \\
 \ldots,  F_{n1}(X_{(\ell+p-1),1})- F_{n1}(y_{p1}),\ldots, F_{nd}(X_{(\ell+p-1),d})- F_{nd}(y_{pd}) \}.
\end{multline*}
Similarly, for all  $x_j\in \RR$ and $\yf_{j}^\ast\in\RR^{p}$, the
conditional marginal cdf's $m_j(x_j\ |\ \yf_j^\ast)$ is estimated
in a nonparametric way by
\begin{multline*}
m_{n,j}(x_j\ |\ \yf_j^\ast)= \frac{1}{n-p}\sum_{\ell=1}^{n-p} \bar K_{\bar h}
\{ F_{nj}(X_{\ell,j})- F_{nj}(y_{1j}),\ldots,\\
F_{nj}(X_{\ell+p-1,j})- F_{nj}(y_{pj}) \} \1 ( \Xf_{\ell+p,j} \leq
x_j),
 \end{multline*}
for every $j=1,\ldots,d$.~\cite{JDF_FermWeg} proposed to estimate
the underlying conditional pseudo-copula by
\begin{equation*}
\widehat C(\uf\ |\ \Xf_{n-1}^{n-p}=\yf^\ast)=m_n\{
m_{n,1}^{(-1)}(u_1\ |\ \yf_1^\ast),\ldots, m_{n,d}^{(-1)}(u_d\ |\
\yf_d^\ast)\ |\ \yf^\ast \}, \label{estimate}
\end{equation*}
with the use of pseudo-inverse functions. Then, under ${\mathcal
H}_0 ^{(1)}$, for all $\uf\in [0,1]^d$ and
$\yf^\ast=(\yf_1,\ldots,\yf_p)\in\RR^{dp} $,
 $$ \sqrt{nh_n^{pd}} \, \{ \widehat C(\uf\ |\
\Xf_{n-1}^{n-p}=\yf^\ast)-C_0(\uf) \} \stackrel{\rm d}{
\mathop{\longrightarrow} \limits_{} } {\mathcal N} [0,
\sigma(\uf)] $$ as $n \to \infty$, where
$\sigma(\uf)=C_0(\uf)\{1-C_0(\uf)\}\int K^2 ({\mathbf v}) \,
\mbox{\rm d} {\mathbf v}.$ This result can be extended to deal
with different vectors $\yf^\ast$ simultaneously, and with the
null hypotheses ${\mathcal H}_0^{(2)}$ and ${\mathcal H}_0^{(3)}$:
for all $\uf\in\RR^d$,
\begin{multline*}
\sqrt{nh_n^{pd}} \, \{ \widehat C(\uf\ |\
\yf_1^\ast)-C_{\hat\theta_1}(\uf),\ldots,\widehat C(\uf\ |\
\yf_q^\ast)-C_{\hat\theta_q}(\uf)\} \stackrel{\rm d}{
\mathop{\longrightarrow}\limits_{} } {\mathcal N}[0,
\Sigma(\uf,\yf_1^\ast,\ldots,\yf_q^\ast)],
\end{multline*}
as $n \to \infty$, where
$$\Sigma(\uf,\yf_1^\ast,\ldots,\yf_q^\ast)=\text{\rm diag} \left(C_{\theta(\yf_k^\ast)}(\uf)
\{1-C_{\theta(\yf_k^\ast)}(\uf)\}\int K^2 ({\mathbf v}) \,
\mbox{d} {\mathbf v}, \ 1\le k\le q \right),
 $$
for some consistent estimators $\hat\theta_k$ such that
$\hat\theta_k=\theta(\yf_k^\ast)+ O_P(n^{-1/2}),\ k=1,\ldots,q$.

Each $k$th term on the diagonal of $\Sigma$ can be consistently
estimated by
 $$
\hat \sigma_k^2(\uf) =  C_{\hat\theta_k}(\uf)
\{1-C_{\hat\theta_k}(\uf) \}\int K^2 ({\mathbf v}) \, \mbox{d}
{\mathbf v}.
 $$

Note that, in the corollary above, the limiting correlation matrix
is diagonal because we are considering different conditioning values
$\yf_1^\ast,\ldots, \yf^\ast_q$ but the same argument $\uf$. At the
opposite, an identical conditioning event but different arguments
$\uf_1,\uf_2,\ldots$ would lead to a complex (non diagonal)
correlation matrix, as explained in~\cite{JDF_Ferm}. The latter weak
convergence result of random vectors allows the building of GOF
tests as in section~\ref{JDF_sec:2}. For
instance, as in~\cite{JDF_Ferm}, a simple test procedure may be
 $$ T(\uf,\yf_1^\ast,\ldots,\yf_q^\ast) = (n h_n^{pd})\sum_{k=1}^q
\frac{ \{ \widehat C(\uf\ |\
\Xf_{n-1}^{n-p}=\yf_k^\ast)-C_{\hat\theta_k}(\uf) \}^{2}
}{\hat\sigma^2_{\yf_k^\ast}(\uf)},
 $$
for different choices of $\uf$ and conditioning values
$\yf_k^\ast$. Under ${\mathcal H}_0 ^{(1)}$, the term on the
right-hand-side tends to a $\chi^2(q)$ distribution under the null
hypothesis. Note that this test is ``local'' since it depends
strongly on the choice of a single $\uf$. An interesting extension
would be to build a ``global'' test, based on the behavior of the
full process
 $$
\sqrt{n h_n^{pd}} \, \{ \widehat C(\cdot \ |\
\Xf_{n-1}^{n-p}=\yf_k^\ast)-C_{\hat\theta_k}(\cdot) \}.
 $$
But the task of getting pivotal limiting laws is far from easy, as
illustrated in~\cite{JDF_Ferm}.

\medskip

In practice, authors often restrict themselves to the case of
time-dependent copula parameters instead of managing time-dependent
multivariate cdfs' nonparametrically. For instance, every
conditional copula or pseudo-copula is assumed to belong to the
Clayton family, and their random parameters $\theta$ depend on the
past observations.~\cite{JDF_Abegaz} has proposed a non-parametric
estimate $\hat\theta(\cdot)$ of the function $\theta$, in the case
of a univariate conditioning variable. It seems possible to build
some GOF tests based on this estimate and its limiting behavior, at
least for simple null hypothesis, but the theory requires more
developments.

\medskip

\section{Practical performances of GOF copula tests}\label{JDF_sec:6}

Once a paper introduces one or several new copula GOF tests, it is
rather usual to include an illustrative section. Typically, two
characteristics are of interest for some tests in competition: their
ability to maintain the theoretical levels powers, and their power
performances under several alternatives. Nonetheless, these
empirical elements, even useful, are often partial and insufficient
to found a clear judgement. Actually, only a few papers have studied
and compared the performances of the main previous tests in depth.
Indeed, the calculation power required for such a large analysis is
significant. That is why a lot of simulation studies restrict
themselves to bivariate copulas and small or moderate sample sizes
(from $n=50$ to $n=500$, typically). The most extensive studies of
finite sample performances are probably those of~\cite{JDF_Berg}
and~\cite{JDF_BeauGenestRem}. In both papers, the set of tests under
scrutiny contains the three main approaches:
\begin{enumerate}
\item
``brute-force'' proposals like $T_n^{KS}$ and/or $T_n^{CvM}$, as
in section~\ref{JDF_sec:2};
\item Kendall's process based tests;
\item
test statistics invoking the PIT (see section~\ref{JDF_sec:3}).
\end{enumerate}
These works found that a lot of tests perform rather well, even
for small samples (from $n=50$, e.g.). Moreover, it is difficult
to exhibit clear hierarchy among all of these tests in terms of
power performances. As pointed out by~\cite{JDF_BeauGenestRem},
\begin{quotation}
No single test is preferable to all others, irrespective of the
circumstances.
\end{quotation}

\medskip

In their experiments,~\cite{JDF_BeauGenestRem} restricted
themselves to bivariate copulas and small sample sizes $n \in
\{50,150\}$. The statistics based on Kendall's dependence function
are promoted, particularly when the underlying copula is assumed
to be Archimedean. It appeared that Cramer-von-Mises style test
statistics are preferable to Kolmogorov-Smirnov ones, all other
things being equal, and whatever the possible transformations of
the data and/or the reductions of information. Among the tests
based on a Cramer-von-Mises statistic, it is difficult to
discriminate between the three main approaches.

\medskip

The latter fact is confirmed in~\cite{JDF_Berg}, that led some
simulated experiments with higher dimensions $d\in \{2,4,8\}$ and
larger sample sizes $n\in \{100,500\}$.~\cite{JDF_Berg} observed the
particularly good performances of a new test statistic, calculated
as the average of the three approaches. Moreover, he studied to
impact of the variables ordering in the PIT. Even if estimated
$p$-values may be different, depending on which permutation order is
chosen, this does not seem to create worrying discrepancies.

\medskip

Notably~\cite{JDF_BergQuessy} led an extensive simulated
experiment of the same type, but their main focus was related to
detecting small departures from the null hypothesis. Thus, they
studied the asymptotic behavior of some GOF test statistics under
sequences of alternatives of the type
$$\H_{a,n}: C= (1-\delta_n)C_0+\delta_n D,$$ where
$\delta_n=n^{-1/2}\delta$, $\delta >0$, and $D$ is another copula. They computed local power curves and compared them for
different test statistics. They showed that the estimation strategy
can have a significant impact on the power of Cramer-von-Mises
statistics and that some ``moment-based'' statistics provide very
powerful tests under many distributional scenarios.

\medskip

Despite the number of available tests in the literature, the
usefulness of all these procedures in practice has to be proved more
convincingly. Apparently, some authors have raised doubts about the
latter point. For instance,~\cite{JDF_Weiss2011a} has evaluated the
performances of Value-at-Risk or VaR (quantiles of loss) and
Expected Shortfall or ES (average losses above a VaR level)
forecasts, for a large set of portfolios of two financial assets and
different copula models. They estimate static copula models on
couples of asset return residuals, once GARCH(1,1) dynamics have
been fitted for every asset independently. They applied three
families of GOF tests (empirical copula process, PIT, Kendall's
function) and five copula models. They found that,
\begin{quotation}
Although copula models with GARCH margins yield considerably better
estimates than correlation-based models, the identification of the
optimal parametric copula form is a serious unsolved problem.
\end{quotation}
Indeed, none of the GOF tests is able to select the copula family
that yields the best VaR- or ES-forecasts. This points out the
difficulty of finding relevant and stable multivariate
dynamics models, especially related to joint extreme moves.
But, such results highlight the fact that it remains a significant
the gap between good performances with simulated experiments and
trustworthy multivariate models, even validated formally by
statistical tests.

\medskip

Indeed, contrary to studies based on simulated samples drawn from an
assumed copula family (the standard case, as in~\cite{JDF_BeauGenestRem}
or~\cite{JDF_Berg}), real data can suffer from outliers or
measurement errors. This is magnified by the fact that most
realistic copulas are actually time-dependent
(\cite{JDF_Weiss2011a}) and/or are mixtures or copulas
(\cite{JDF_KoleEtAl}). Therefore,~\cite{JDF_Weiss2011b} showed that
even minor contamination of a dataset can lead to significant power
decreases of copula GOF tests. He applied several outlier detection
methods from the theory of robust statistics, as
in~\cite{JDF_Mendes}, before leading the formal GOF test of any
parametric copula family.~\cite{JDF_Weiss2011b} concluded that the
exclusion of outliers can have a beneficial effect on the power of
copula  GOF tests.

\end{document}